\documentclass[%
reprint,
preprintnumbers,
nofootinbib,
floatfix,
amsmath,amssymb,
prstab,
longbibliography
]{revtex4-2}

\newcommand{\imgpath}{.}

\newcommand{\includetikz}[1]{%
	\includegraphics{#1.pdf}%	
}
\newcommand{\includetikzrename}[2]{%
	\includegraphics{#2.pdf}%	
}

\usepackage{dcolumn}% Align table columns on decimal point
\usepackage{graphicx}

\usepackage{siunitx}
\usepackage{booktabs}
\usepackage{enumerate}
\usepackage{standalone}
\usepackage{mathtools}
\usepackage{bm}

\usepackage{amsthm}
\newtheoremstyle{remarkstyle} % name
{}                    % Space above
{}                    % Space below
{}                   % Body font
{}                           % Indent amount
{\itshape}                   % Theorem head font
{}                          % Punctuation after theorem head
{.5em}                       % Space after theorem head
{\thmname{#1} \thmnumber{#2}\thmnote{, #3}: }  % Theorem head spec (can be left empty, meaning ‘normal’)

\theoremstyle{remarkstyle}
\newtheorem{remark}{Remark}

\usepackage{commands}

\usepackage{balance}

\renewcommand{\fg}{\mathbf{f}_{\mathbf{g}}}

\newcommand{\Avec}{\underline{\bA}}

\renewcommand{\toprule}{\midrule\midrule}
\renewcommand{\bottomrule}{\midrule\midrule}

\let\revappendix\appendix % cleveref overwrites the starred appendix version
\usepackage[capitalise]{cleveref}

\usepackage{microtype}

\begin{document}

\preprint{}%APS/123-QED}

\title{Modeling same-order modes of multicell cavities}

\author{Olof Troeng}
\thanks{E-mail: \texttt{oloft@control.lth.se}}%

\affiliation{%
	Department of Automatic Control, Lund University, Sweden
}

\date{\today}% It is always \today, today,

\begin{abstract}
We derive the transfer function of a multicell cavity with parasitic same-order modes (from power coupler to pickup probe).
The derived model is discussed and compared to measurement data.
\end{abstract}

%\pacs{DRAFT}% PACS, the Physics and Astronomy
\maketitle

\section{Introduction}
Multicell elliptical cavities are suitable for accelerating particle beams with velocities greater than $0.5c$.
However, these cavities intrinsically have parasitic electromagnetic modes that are close in frequency to the accelerating mode.
These so-called \emph{same-order modes}, or fundamental passband modes, interact both with the \rf{} system and the beam.
Their interaction with the \rf{} system may, without counter-measures, give instability in the field control loop \cite{Schilcher1998}, and their interaction with the beam drives emittance growth \cite{Ainsworth2012}.
Dynamic models of same-order modes and these interactions are necessary for design and analysis of field control algorithms.

The regular geometry of multicell elliptical cavities makes it possible to derive the same-order-mode dynamics from a small number of cavity parameters. Such dynamic models have been used for studying how beating of same-order modes affects the  beam  \cite{Ferrario1996}.
The steady-state models in \cite{Doolittle1989,Padamsee2008,Wangler2008,Sekutowicz2010} capture the shapes and resonance frequencies of the modes but not their dynamics and decay rates.

For field control analysis in the frequency domain, it is necessary to know the  transfer function from the \rf{} drive to the pickup-probe signal.
In \cite{Schilcher1998}, a real-coefficient, two-input two-output transfer function from \rf{} drive to pickup signal (valid for superconducting cavities) was presented without motivation. 
Such real-coefficient, two-input two-output models are common in the field control literature. However, as we discussed in \cite{Troeng2019Thesis}, the equivalent, complex-coefficient, single-input single-output (SISO) representation gives more intuition and simplifies analysis. 

In this paper we: (1) derive a complex baseband model of a multicell cavity along the lines of \cite{Ferrario1996}, but using the energy-based parameterization from \cite{Troeng2020CavityModel}; (2) normalize that model to make it suitable for field control analysis; (3) derive the complex-coefficient SISO transfer function from power coupler to pickup signal (also valid for normal-conducting cavities); and (4) fit the transfer-function model to measurement data from a 6-cell niobium cavity taken at room temperature and cryogenic temperature.

\begin{remark}
An incorrect transfer-function model for cavities with parasitic modes was proposed in \cite{Vogel2007}.
It was based on taking the transfer function of the individual modes from \cite{Schilcher1998} and multiplying them with the modes' coupling strengths to the \rf{} system, although that effect was already accounted for in \cite{Schilcher1998}. The incorrect model in \cite{Vogel2007} predicts  that the resonance peaks have different magnitudes, while both the analysis in this paper and \cite{Liepe2001}[Fig.~5.9] indicate that they have similar magnitudes.

Also the stability analysis in \cite{Vogel2007} is problematic.
It is not possible to analyze closed-loop stability of MIMO systems with strong cross couplings (as from same-order modes) using loop-by-loop analysis \cite[Sec. 8.6]{Zhou1996}, \cite{Skogestad2007}.
The complex-signal perspective in \cite{Troeng2019Thesis} together with the model in this paper enables correct and intuitive analysis.
\end{remark}

\section{Same-Order Modes of multicell Cavities}
\label{sec:parasitic_modes}

\begin{figure}
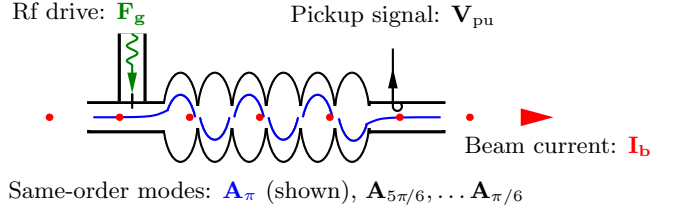

	\centering
	\includetikz{\imgpath/elliptical_cavity_graphic}
	\caption{Schematic of 6-cell elliptical cavity.
	Power is fed through the power coupler in the leftmost cell and the cavity field is sensed by a pickup probe in the rightmost cell.}
	\label{fig:elliptical_cavity_graphic}
\end{figure}

\subsection{Cavity model with parasitic modes}
We start by considering parasitic modes of a general cavity using the energy-based parameterization in \cite{Haus1983,Troeng2020CavityModel}. Let $\bA_k$ be the complex envelope of mode $k$, with its magnitude (units~\si{\sqrt{J}}) scaled so that $\abs{\bA_k}^2$ is the mode energy. This unambiguously defines the mode amplitude without reference to the effective voltage of the accelerating mode.
If coupling between the modes can be neglected then their complex envelopes evolve according to \cite{Troeng2020CavityModel}
\begin{subequations}
\begin{equation}
\d{\bA_k}{t}  = (-\gamma_k + i\Dw_k) \bA_k  + \sqrt{2{\gamma_\ext}_k} \Fg + \frac{\bm{\alpha}_k}{2}  \Ib,
\label{eq:mode_dynamics}
\end{equation}
where $\gamma_k = {\gamma_0}_k + {\gamma_\ext}_k$ is the decay rate of the mode amplitudes (${\gamma_0}_k$ corresponds to resistive losses and ${\gamma_\ext}_k$  to decay through the power coupler); $\Dw_k = \omega_k - \wRF$ is the offset of the mode frequency $\omega_k$ relative to the nominal \rf{} frequency $\wRF$; $\Fg$ (units \si{\sqrt{W}}) is the envelope of the forward wave from the \rf{} amplifier with $\abs{\Fg}^2$ equal to the power in the wave; and $\Ib$ is the beam phasor, with $\abs{\Ib}$ equal to the dc beam current. 
The complex envelopes $\bA_k$ are defined relative to $\wRF$, with their phases chosen so that the coefficients	 in front of $\Fg$ are real.
The voltage sensed by the pickup probe is a linear combination
\begin{equation}
\Vpu = \sum_{k=a,1,2,\ldots}^N \bm{C}_k \bA_k
\label{eq:pickup_lincomb}
\end{equation}
\label{eq:before_normalization}%
\end{subequations}
of the mode amplitudes, where $\bm{C}_k$ are complex coefficients.
Here we have used the subscript $a$ to indicate the mode intended for acceleration. We will use a different labeling when discussing same-order modes.

The cavity--beam-coupling parameters $\bm{\alpha}_k$ are in general complex. However, the parameter \mbox{$\bm{\alpha}_a = \alpha_a$} for the accelerating mode is real due to the definition of $\Ib$.

\subsection*{Normalized model with parasitic modes}
In many situations, e.g., field control analysis, it is easier to work with normalized models.
Therefore, we introduce the following dimension-free, normalized variables similarly to in \cite{Troeng2020CavityModel},
\begin{subequations}
	\begin{align}
	\ba_k &\coloneqq \,\, \frac{1}{A_{a0}} \, \bA_k \\[0.2em]
	\fg &\coloneqq \frac{1}{\gamma_a A_{a0}}  \sqrt{2{\gamma_\ext}_a}\Fg \\[0.2em]
	\ib &\coloneqq \frac{1}{\gamma_a A_{a0}} \frac{\alpha_a}{2} \Ib\\
	\vpu &\coloneqq \frac{1}{C_a A_{a0}} \Vpu,
	\end{align}%
	\label{eq:parasitics_normalized_parameters}%
\end{subequations}
where $A_{a0}$ is the nominal magnitude of the accelerating mode.
With the variables \eqref{eq:parasitics_normalized_parameters} we can write \eqref{eq:before_normalization} as
\begin{subequations}
	\begin{align}
	\d{\ba_k}{t}  &= (-\gamma_k + i\Dw_k) \ba_k  + \gamma_a \frac{\sqrt{2{\gamma_\ext}_k}}{\sqrt{2{\gamma_\ext}_a}} \fg + \gamma_a \frac{\bm{\alpha}_k}{\alpha_a}  \ib. \\
	\vpu &= \ba_a + \sum_{k=1,2,\ldots}^N  \bm{c}_k \ba_k
	\end{align}
	\label{eq:after_normalization}%
\end{subequations}
where $\bm{c}_k \coloneqq \bm{C}_k / (C_a A_{a0})$.
The normalization \eqref{eq:after_normalization} has the nominal operating point $\ba_a = 1$ and the steady-state sensitivity of $\ba_a$ to variations in $\fg$ and $\ib$ is unity.

\subsection{Same-order modes of multicell elliptical cavities}
\label{sec:parasitics_elliptical}

The same-order modes of a multicell cavity arise from the coupling between the cells' fundamental modes, similarly as for a chain of weakly coupled oscillators.
An $N$-cell cavity has $N$ closely spaced same-order modes. 
The parameters in \eqref{eq:after_normalization} of these modes, can, due to the regular cavity geometry, be computed from a small number of basic cavity parameters, as shown in the Appendix.
The modes are conventionally referred to as the $\pi/N$ mode, the $2\pi/N$ mode, up to the $\pi$ mode. This naming indicates the cell-to-cell phase advance \emph{of the sinusoidal envelope of the mode shapes}%
\footnote{\scalebox{0.96}{The cell-to-cell phase difference of the modes themselves is $0$ or $\pi$.}}.
It is typically the $\pi$ mode that is used as the accelerating mode \cite{Padamsee2008}.
We will use a subscript $\pi$ to indicate parameters of the $\pi$~mode and a subscript $n$ for the $n\pi/N$ mode.

\begin{figure}
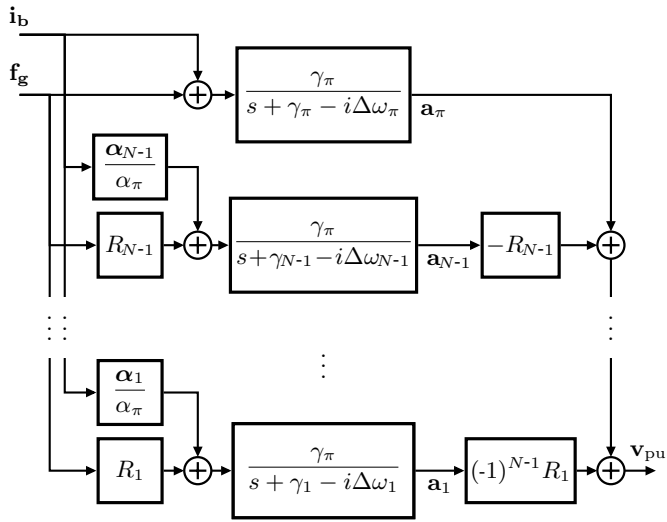

	\centering
	\makebox[0pt][c]{
	\hspace*{-3mm}
	\includetikz{\imgpath/soms_normalized_full}
	}
	\caption{Block diagram for a normalized model of a cavity with parasitic same-order modes. Subscripts $n$ indicate the $n\pi/N$ mode.}
	\label{fig:soms_normalized_full}
\end{figure}

The derivation in the appendix assumes an ``ideal'' \mbox{$N$-cell} cavity in that: all cell-to-cell coupling factors equal $\kcc$; all inner-cells have resonance frequencies equal to $\wcell$; the end-cell resonance frequencies are tuned for a flat $\pi$ mode; and mode coupling is negligible.
Under these assumptions, the same-order-mode parameters depend, in addition to  $N$, $\wcell$, and $\kcc$, only on the cell's resistive decay rate $\gamma_0$ and the decay rate $\gamma_\text{pc}$ through the power coupler from the connected end cell. By introducing
\begin{equation}
R_n \coloneqq 
\begin{cases}
\sqrt{2} \sin \dfrac{n\pi}{2N} &\text{if\,\,\,} n < N\\[0.5em]
1 & \text{if\,\,\,} n = N
\end{cases},
\label{eq:def_Rn}
\end{equation}
the parameters can be expressed as
\begin{align}
\Dw_n & = \wcell \sqrt{1 + 2 R_n^2 \kcc} - \wRF \qquad\quad (n < N) \label{eq:soms_Dw} \\
& \approx \wcell (\sqrt{1 + 2 R_n^2 \kcc} - \sqrt{1 + 4\kcc}) \label{eq:soms_Dw_approx} \tag{\ref{eq:soms_Dw}$^\prime$} \\
{\gamma_0}_n &= \gamma_0 \label{eq:soms_gamma0} \\
{\gamma_\ext}_n &= R_n^2 \gamma_\text{pc}/N = R_n^2 {\gamma_\ext}_\pi \label{eq:soms_gammaext}\\
\bm{c}_n &= (-1)^{N-n} R_n \label{eq:soms_cn}.
\end{align}
The approximation \eqref{eq:soms_Dw_approx} assumes that $\Dw_\pi = \omega_\pi - \wRF$ is small relative to $\Dw_{N-1}$, see also \cref{rmk:detuning}.
With the relationships \eqref{eq:soms_Dw}--\eqref{eq:soms_cn} the general model in \eqref{eq:after_normalization} takes the form in \cref{fig:soms_normalized_full}. Values of $R_n^2$ for different values of $N$ and $n$ are shown in \cref{tab:Rn_values}. In the next section we investigate how well the relations \eqref{eq:soms_Dw}--\eqref{eq:soms_cn} agree for parameters estimated from a real-world cavity.

\begin{table}
	\centering
	\setlength{\tabcolsep}{4pt}
	\caption{Values of $R_n^2$ for different $n$ and $N$, see  \eqref{eq:def_Rn} for $R_n$.}
	\label{tab:Rn_values}
	\vspace{0.5em}
	\begin{tabular}{l@{\hskip 20pt}cccccccccc}
		\toprule
		$N$ & \multicolumn{8}{c}{$n$} \\
		\cmidrule{2-10}
		& 9 & 8 & 7 & 6 & 5 & 4 & 3 & 2 & 1 \\
		\midrule
		5 & & & & & 1 & 1.81 & 1.31 & 0.69 & 0.19  \\
		6 & & & & 1 & 1.87 & 1.50 & 1.00 & 0.50 & 0.13  \\
		9 & 1 & 1.94 & 1.77 & 1.50 & 1.17 & 0.83 & 0.50 & 0.23 & 0.06  \\
		\bottomrule
	\end{tabular}
\end{table}

\begin{remark}
\label{rmk:detuning}
While $\Dw_\pi$ is negligible relative to both $\Dw_{N-1}$ and typical field control bandwidths, its precise tuning, typically to a value slightly larger than 0, is crucial for minimizing the drive power $\abs{\Fg}^2$ \cite{Wangler2008,Padamsee2008,Troeng2020CavityModel}.
\end{remark}
\begin{remark}
Typical cell-to-cell coupling factors $\kcc$ are on the order of $0.01$, which gives
\begin{align*}
\Dw_n & \approx \wcell \left( \sqrt{1 + 2R_n^2 \kcc} - \sqrt{1 + 4\kcc} \right) \\
&\approx (R_n^2 - 2)\kcc \wcell.
\end{align*}
This shows that the baseband frequencies of the same-order modes are approximately proportional to $R_n^2 - 2$.
\end{remark}
\begin{remark}
	There are two definitions of the cell-to-cell coupling factor $\kcc$ in the accelerator literature.
	The \emph{full-passband-width} definition, for which $\kcc \approx (\omega_\pi  - \omega_{1})/\omega_\pi$   \cite{Doolittle1989,Liepe2001,Wangler2008,Sekutowicz2010} and the \emph{half-passband-width} definition \cite{Ferrario1996,Padamsee2008} for which $\kcc$ is half as large.
	The first definition corresponds to the per-cycle decay of a cell's energy due to coupling to a neighboring cell and the latter definition corresponds to the decay of the field amplitude.
	In this paper we use the half-passband-width definition since it gives slightly more convenient expressions.
\end{remark}

\subsection{\!\!Transfer function from \rf{} drive to pickup probe\!}
The transfer function from $\fg$ to $\vpu$ in \cref{fig:soms_normalized_full} is important for field control analysis. It is given by
\begin{align}
P_\text{cav}(s)
&= \gamma_\pi \sum_{n=1}^N (-1)^{N-n} \frac{R_n^2}{s + \gamma_n - i\Dw_n}
\notag
\\
&= \frac{\gamma_\pi}{{\gamma_\ext}_\pi} \sum_{n=1}^N (-1)^{N-n} \frac{{\gamma_\ext}_n}{s + \gamma_n - i\Dw_n},
\label{eq:P_cav_elliptical}
\end{align}
%\end{subequations}
where $\gamma_n \coloneqq \gamma_0 + {\gamma_\ext}_n$ is  the total decay rate of the $n\pi/N$ mode. For superconducting cavities with $\gamma_0 \ll {\gamma_\ext}_n$ we have that  ${\gamma_\ext}_n \approx \gamma_n$ and \eqref{eq:P_cav_elliptical} simplifies to
\begin{equation}
P_\text{cav}(s)
=  \sum_{n=1}^N (-1)^{N-n} \frac{\gamma_n}{s + \gamma_n - i\Dw_n}.
\label{eq:P_cav_elliptical_sc}
\end{equation}

Let us observe some characteristics of the transfer function \eqref{eq:P_cav_elliptical}.
First, since the numbers $\gamma_n$ are small relative to the differences between the numbers $\Dw_n$, the transfer function $\Pcav(s)$ has sharp resonance peaks at (baseband) frequencies $\Dw_n$.
Also note that $\Dw_n < 0$ for $n < N$, i.e., the baseband resonance frequencies of all parasitic same-order modes are negative.

For superconducting cavities ($\gamma_0 \ll {\gamma_\ext}_n$)  we see from \eqref{eq:P_cav_elliptical_sc} that all peaks have approximately unity magnitude.
For normal conducting cavities ($\gamma_0 \gg {\gamma_\ext}_n$) we have that the peak magnitude of the $n\pi/N$ mode equals ${\gamma_\ext}_n/{\gamma_\ext}_\pi = R_n^2$ (see \cref{tab:Rn_values}).
\cref{fig:cavity_soms_bode} shows the Bode magnitude plots of these extreme cases.

\section{Comparison to measurement data}
\label{sec:comparison_measurement_data}

In this section we examine the fit of relationships \eqref{eq:soms_Dw}--\eqref{eq:P_cav_elliptical} to network-analyzer measurements of a 6-cell medium-$\beta$ cavity%
\footnote{The cavity was a prototype without a tuning system, hence the mode frequencies during the measurements differed from the design frequencies.
For the $\pi$-mode,  which should have a nominal frequency of \SI{704.42}{MHz}, the measured frequencies were \SI{703.26}{MHz} (NC) and   \SI{704.24}{MHz} (SC).
That the resonance frequency is significantly lower at room temperature is typical.
}
for the European Spallation Source (ESS).
Measurements were taken when the cavity was normal conducting (NC, at room temperature) and superconducting (SC, at \SI{2}{K}).
In the first case  $\gamma_0 \gg {\gamma_\ext}_n$ and in the latter case $\gamma_0 \approx 0$.
The measurements (scaled and frequency shifted for a unity-gain $\pi$ mode at zero frequency) are shown in gray in \cref{fig:med_beta_data_fit}.

The observed same-order-mode frequencies (relative to the $\pi$ mode) are shown on the frequency axes in \cref{fig:med_beta_data_fit}.
From those values, the cell-to-cell coupling factor (half-passband-width definition) was estimated to $\kcc \approx 0.00862$, which makes the expression for $\Dw_n$ in \eqref{eq:soms_Dw} agree with the observed frequency offsets to within $\pm 1 \%$.

\begin{figure}
	\centering	
	{
	\newcommand{\sccavity}{}
	\includetikzrename{\imgpath/soms_datafit_semilogy}{\imgpath/soms_data_fit_sc}
	
	{(a) \footnotesize Superconducting (\SI{2}{K}).}
	}
	\includetikzrename{\imgpath/soms_datafit_semilogy}{\imgpath/soms_data_fit_nc}
	
	{(b) \footnotesize Normal conducting (room temperature).}
	
	\caption{Network-analyzer measurements of the transmission of a 6-cell ESS medium-$\beta$ cavity (gray) and fits of the transfer function \eqref{eq:P_cav_elliptical} (blue, orange).
	The data is scaled for unity gain of the $\pi$ mode. Measurements by P. Pierini, ESS.}
	\label{fig:med_beta_data_fit}
\end{figure}

From the measurement data and the observed frequency offsets $\Dw_n$, the remaining parameters in \eqref{eq:soms_Dw}--\eqref{eq:soms_cn}, namely $\gamma_0$ and ${\gamma_\ext}_\pi$, were estimated by fitting \eqref{eq:P_cav_elliptical} to the data. A good fit was obtained with ${\gamma_\ext}_\pi/2\pi = \SI{460}{Hz}$, $\gamma_0 = 0$ (SC), and $\gamma_0 / 2\pi = \SI{35}{kHz}$ (NC). \Cref{fig:med_beta_data_fit} shows the fitted models and the measurement data.

For comparison, we estimated $\gamma_k$ and the peak magnitude $g_k$ for each individual mode by fitting $g_k\gamma_k/(i(\omega-\omega_k) + \gamma_k)$ to the data in the vicinity of each mode. In Table~\ref{tab:mode_parameters}, these ``observed'' mode parameters are compared to those predicted by \eqref{eq:soms_gamma0}, \eqref{eq:soms_gammaext}, and \eqref{eq:P_cav_elliptical}.
It is seen that they agree reasonably well---in particular for the modes closest to the $\pi$ mode, which are the most crucial ones in a field control context.
The small discrepancies are probably explained by variations in the cell parameters.

\begin{table}
	\newcommand{\mmp}[2]{\num{#1} & {\scriptsize($#2$\%)}}
	\centering
	\setlength{\tabcolsep}{4pt}
	\caption{Comparison between peak magnitudes and bandwidths for measurement data and model fits in \cref{fig:med_beta_data_fit}. The observed values are shown and the value in parenthesis indicates the deviation from the value predicted by the model.}
	\label{tab:mode_parameters}
	\begin{tabular}{@{}cr@{\hspace{1mm}}lr@{\hspace{1mm}}l@{\hspace{4mm}}r@{\hspace{1mm}}lr@{\hspace{1mm}}l@{}}
		\toprule	
		& \multicolumn{4}{c}{Superconudcting}
		& \multicolumn{4}{c}{Normal conducting}
		\\
		\cmidrule(lr){2-5} \cmidrule{6-9}
		& \multicolumn{2}{c}{\!$\abs{\Pcav(i\Dw_k)}$\!}
		& \multicolumn{2}{c}{$\gamma_k/2\pi$}
		& \multicolumn{2}{c}{\!$\abs{\Pcav(i\Dw_k)}$\!}
		& \multicolumn{2}{c}{$\gamma_k/2\pi$}
		\\
		& \multicolumn{2}{c}{}
		& \multicolumn{2}{c}{\sibrac{Hz}}
		& \multicolumn{2}{c}{}
		& \multicolumn{2}{c}{\sibrac{kHz}}
		\\
		\midrule
		$\pi$  & 1.00 & & \mmp{450}{1.6} & 1.00 & & \mmp{36}{-2} \\ 
		$ 5\pi/6$ &\mmp{1.03}{-3} &\mmp{870}{-2} &\mmp{1.85}{-0.2} &\mmp{35}{3} \\ 
		$ 4\pi/6$ &\mmp{1.01}{-1} &\mmp{770}{-11} &\mmp{1.49}{0.3} &\mmp{35}{3} \\ 
		$ 3\pi/6$ &\mmp{1.14}{-13} &\mmp{500}{-9} &\mmp{1.07}{-6} &\mmp{34}{3} \\ 
		$ 2\pi/6$ &\mmp{1.12}{-11} &\mmp{270}{-14} &\mmp{0.56}{-10} &\mmp{34}{5} \\ 
		$ \pi/6$ &\mmp{0.93}{7} &\mmp{72}{-14} &\mmp{0.14}{-5} &\mmp{32}{9} \\		
		\bottomrule
	\end{tabular}
\end{table}

\begin{figure}
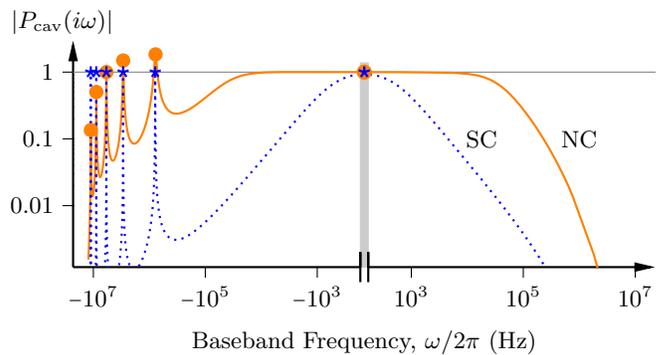

	\centering
	\includetikz{\imgpath/soms_dloglog_comparison}	
	\caption{Bode magnitude plots of the transfer function \eqref{eq:P_cav_elliptical} for a 6-cell cavity when it is superconducting (SC, $\gamma_0 = 0$) and normal conducting (NC, $\gamma_0 / 2\pi= \SI{35}{kHz}$). Note that the transfer functions have been scaled for unity magnitude at the zero frequency, see Remark~\ref{rmk:trans_scaling}.}
	\label{fig:cavity_soms_bode}
\end{figure}

For field control analysis, it is convenient to plot frequency responses using a logarithmic frequency axis, in so-called Bode diagrams.
The frequency response of a normal conducting and a superconducting cavity are shown in a double-sided Bode diagram in \cref{fig:cavity_soms_bode}. 
Note that superconducting cavities, whose external decay rates ${\gamma_\ext}_\pi$ are significantly larger than $\gamma_0$, have resonance peaks with approximately unity magnitude (\cref{fig:cavity_soms_bode}).
For normal conducting cavities (with small ${\gamma_\ext}_\pi$), the peak magnitudes are approximately given by $R_n^2$.

\begin{remark}
\label{rmk:trans_scaling}
Recall that the data in \cref{fig:med_beta_data_fit} and \cref{fig:cavity_soms_bode} is scaled for unity magnitude of the $\pi$ mode. In absolute terms, the transmission of the normal conducting cavity is lower than for the superconducting cavity. 
\end{remark}

\section{Conclusion}
We have derived a complex-coefficient transfer function model that is valid for both normal conducting and superconducting cavities. The model was seen to give a good fit to measurement data.

\begin{acknowledgments}
Bo Bernhardsson and Paolo Pierini contriubted helpful comments and suggestions. %Daniel Sjöberg,  Larry Doolittle, Lars Rippe, and  for helpful comments and suggestions.
The measurement data in Sec.~\ref{sec:comparison_measurement_data} was provided by Paolo Pierini.
The author is a member of the ELLIIT Strategic Research Area at Lund University.
\end{acknowledgments}

\revappendix*
\section{Derivation}\label{app:derivation}
In this appendix, we will start from the bandpass state-space model of an $N$-cell cavity in \cite{Ferrario1996}, perform modal decomposition (diagonalization), and then transform the diagonal model to baseband.

\subsection{Bandpass model of an $\bm{N}$-cell cavity}
Our starting point is the standard model for studying same-order modes of multicell cavities \cite{Ferrario1996,Padamsee2008,Sekutowicz2010}, but we will use slightly different notation.
Consider the elliptical $N$-cell cavity in \cref{fig:elliptical_cavity_graphic} which has $N-2$ identical inner cells and two end cells that are joined with the beam pipe.
Cell $1$ can be excited through a power coupler connected to the beam pipe.
Adjacent cells are connected by irises that enable the propagation of the electromagnetic field.
The field is cell $N$ is sensed by a pickup probe mounted in the beam pipe.

We only consider the lowest-energy mode in each cell and we denote the electric field amplitudes of these by $x = \bmat{x_1 & \cdots & x_N}\!\!\phantom{|}^\tran$.
We assume that $x_\ell$ is normalized so that the squared magnitude of its complex envelope equals the energy stored in cell $\ell$.
Let all cell-to-cell coupling factors be given by $\kcc$; the inner-cell resonance frequencies be given by $\omega_\cell$; the end-cell resonance frequencies be given by $\omega_\text{cell} \sqrt{1 + 2\kcc}$; the \rf{} drive (i.e., the forward wave entering the power coupler) be modeled by its complex envelope $\Fg$, with $\abs{\Fg}^2$ equaling the power in the forward wave; the coupling between the waveguide and the field in cell 1 be quantified by the decay rate $\gamma_\text{pc}$ of the field in this through the power coupler; and assume that field decay through the pickup probe is negligible.
According to \cite[(B-1)]{Ferrario1996} the field amplitudes $x$ in the cells evolve as a chain of weakly coupled oscillators, with the dynamics
\begin{multline}
\ddot{x} + 2\gamma_0 \dot{x} + 2\gamma_\text{pc} E_1 \dot{x}
+ \wcell^2 x + \wcell^2  \kcc K x \\
= 2\sqrt{2\gamma_\text{pc}} e_1 \frac{d}{dt}\Re\{ \Fg \me^{i\wRF t} \},
\label{eq:coupled_cells_matrix_form}
\end{multline}
where
\begin{align}
K &=
\bmat{
	3 & - 1	& 0 	 & \cdots 	& 	0 		\\
	- 1 	 & 2 & -1 	 &  0		& 	\vdots 	\\
	0 		 & 0 		& \ddots &  		&	    0 	\\
	\vdots  &  		& -1 	 & 2	& 	-1 	\\
	0		 &  \cdots	& 0		 & -1 	&  3
}, \label{eq:K_matrix}
\\[0.5em]
E_1 &= \text{diag}(1, 0, \ldots, 0), \notag \\[0.3em]
e_1 &= \bmat{1 & 0 & \ldots & 0}^\tran.
\notag
\end{align}

% Grounded Laplacian...

\subsection{Eigenvectors and eigenvalues of the matrix $K$}

\begin{figure}
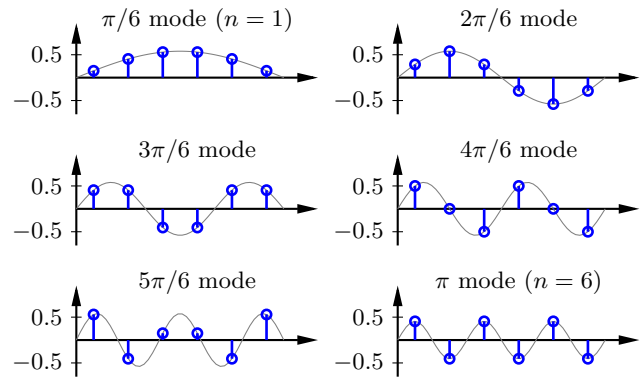

	\centering
	\includetikz{\imgpath/som_shapes}
	
	\caption{Same-order-mode shapes $q_n$ \eqref{eq:som_shapes} of a 6-cell cavity.}
	\label{fig:som_shapes}
\end{figure}

By recalling standard trigonometric identities and doing some algebra---alternatively looking up
\cite[Sec. 7.2]{Padamsee2008}---it can be verified that the matrix $K$ in \eqref{eq:K_matrix} has the eigenfactorization $\bQ \Lambda \bQ^\tran = K$, where
\begin{align}
Q &\coloneqq
\bmat{  |   &   |  &  &  | \\[-1pt]
	q_1 & q_2 & \hdots & q_N \\[1pt]
	|   & 	|	& & |}, \label{eq:Q_matrix}\\
\Lambda &\coloneqq \text{diag}(\lambda_1, \, \lambda_2, \, \cdots, \lambda_N),
\end{align}
and the eigenvalues $\lambda_n$ are given by
\begin{equation}
\lambda_n = 2\left(1- \cos\frac{n\pi}{N} \right),
\label{eq:eigenvalue_soms}
\end{equation}
and the orthonormal eigenvectors $q_n$ are given by
\begin{equation}
\renewcommand*{\arraystretch}{1.3}
\begin{array}{c}
\phantom{N}
\\[0.9em]
\bq_n =
\sqrt{\dfrac{2}{N}}
\\[0.9em]
\text{\small $(n < N)$}
\end{array}
\!\!
\bmat{
	\sin  \left[ (1 - \frac{1}{2}) \frac{n\pi}{N}  \right] \\[0.2em]
	\sin  \left[ (2 - \frac{1}{2}) \frac{n\pi}{N} \right] \\
	\vdots \\
	\sin  \left[ (N - \frac{1}{2}) \frac{n\pi}{N} \right] \\
}
\!,
\quad
\bq_N  =
\sqrt{\frac{1}{N}}
\!
\bmat{
	1 \\[0.2em]
	-1 \\
	\vdots \\[0.2em]
	(-1)^{N-1} \\
}
\!\!.
\label{eq:som_shapes}
\end{equation}
The the mode shapes \eqref{eq:som_shapes} are illustrated in \cref{fig:som_shapes}.
As mentioned in Sec.~\ref{sec:parasitics_elliptical}, mode $n$ is often referred to as the $n\pi/N$ mode.
The entries of the $N$th mode have equal magnitude and opposite signs and for this reason it is almost always used as the accelerating mode.

\subsection{Diagonalizing the dynamics}
Letting $\xi$ be the passband mode amplitudes ($x = \bQ \xi$), we may diagonalize all of \eqref{eq:coupled_cells_matrix_form} except the third term,
\begin{multline}
\ddot{\xi}  + 2\gamma_0 \dot{\xi} + 2 \bQ^\tran E_1 \bQ \gamma_\text{pc}\dot{\xi} +  \omega_\cell^2(I + \kcc\Lambda) \xi \\
= 2\sqrt{2\gamma_\text{pc}} \bQ e_1 \frac{d}{dt} \Re\{ \Fg \me^{i\wRF t} \},
\label{eq:modes_rf}
\end{multline}
where $I$ is the identify matrix.
For convenience, denote $\bQ$'s first row times $\sqrt{N}$ by
\begin{align*}
R \coloneqq &\bmat{\quad\,\, R_1 \qquad\,\,\, \cdots \qquad R_n \qquad \cdots \,\,  R_N}\\
=& \bmat{\sqrt{2}\sin \dfrac{\pi}{2N} & \cdots & \sqrt{2} \sin \dfrac{n\pi}{2N} & \cdots & \,\,1\,\,\,},
\end{align*}
that is $R_n$ is given by \eqref{eq:def_Rn}.
Then we can write the third term of (\ref{eq:modes_rf}) as
$2 R^\tran R / N\gamma_\text{pc}  \dot{\xi}$. This term, which originates from field decay through the power coupler, dynamically couples the modes.
We may however assume that this interaction averages out to $0$, since the beating period between the different modes is significantly shorter than the timescales at which the (complex) mode amplitudes change.
Thus it suffices to keep the diagonal entries of $R^\tran R$, which  correspond to the external decay rates of the modes. Denoting them by
\begin{equation}
{\gamma_\ext}_n \coloneqq R_n^2 \,\gamma_\text{pc}/N,
\label{eq:gamma_ext_n_pre}
\end{equation}
we get $N$ uncoupled differential equations from \eqref{eq:modes_rf}, one for each mode,
\begin{multline}
\ddot{\xi}_n  + 2\gamma_0 \dot{\xi}_n + 2{\gamma_\ext}_n \dot{\xi}_n +  \omega_\cell^2(1 + \kcc\lambda_n) \xi_n \\
= 2 \sqrt{2{\gamma_\ext}_n}  \frac{d}{dt} \Re\{ \Fg \me^{i\wRF t} \}.
\label{eq:modes_rf2}
\end{multline}
From \eqref{eq:gamma_ext_n_pre} we see that the external decay rate of the $\pi$~mode is given by ${\gamma_\ext}_\pi = \gamma_\text{pc}/N$ and hence the external decay rates of the same-order modes satisfy
\begin{equation}
{\gamma_\ext}_{n}
= R_n^2 {\gamma_\ext}_\pi
\label{eq:gamma_ext_n}.
\end{equation}

\subsection{Baseband state-space model}
The eigenfrequencies of the modes in \eqref{eq:modes_rf2} are 
\[
\omega_n = \omega_\cell \sqrt{1 + \kcc \lambda_n}.
\]
See Fig.~7.4 in \cite{Padamsee2008} for an illustration. Denote their offsets from the nominal \rf{} frequency $\wRF$ by $\Dw_n \coloneqq \omega_n - \wRF$.
\iffalse
\begin{figure}
	\centering
	\includetikz{\imgpath/som_frequencies}
	
	\caption{The resonance frequencies of a 6-cell cavity.}
	\label{fig:som_frequencies}
\end{figure}
\fi

Let $\bA_n$ denote the complex envelope of the $n\pi/N$ mode with respect to $\wRF$, i.e., $\xi_n = \Re\{ \bA_n \me^{i\wRF t} \}$. A slowly-varying envelope approximation of \eqref{eq:modes_rf2} is then given by%
\begin{equation}
\dot{\bA}_n =
\left[-(\gamma_0 + {\gamma_\ext}_n) + i \Dw_n \right] \bA_n \\
+ \sqrt{2 {\gamma_\ext}_n} \Fg.
\label{eq:single_mode_evolution_bb}
\end{equation}
By introducing $\Avec \coloneqq \bmat{\bA_1 &\!\cdots\!& \bA_N}^\tran$,  $\Delta\Omega \coloneqq \text{diag}(\Dw_1, \ldots, \Dw_n)$, and $\Gamma_\ext \coloneqq \text{diag}( {\gamma_\ext}_1, \ldots, {\gamma_\ext}_N)$,  we can write the equations \eqref{eq:single_mode_evolution_bb} as
\begin{equation}
\dot{\Avec} =
\left[-(\gamma_0 I + \Gamma_\ext)
+  i \Delta\Omega  \right] \Avec \\
+ \sqrt{2 {\gamma_\ext}_\pi} R^\tran \Fg.
\label{eq:mode_evolution_bb}
\end{equation}

The voltage signal $\Vpu$ from the pickup probe is proportional to the field amplitude in cell $N$, i.e., \mbox{$\Vpu \propto \bQ_{N:} \Avec$}
where $\bQ_{N:}$ denotes the $N$th row of $Q$. Using that%
\iffalse\footnote{$
	Q_{Nn} = \sin  \left[ (N - \frac{1}{2}) \frac{n\pi}{N} \right]
	=
	\sin  \left[ n\pi - \frac{n\pi}{2N} \right]
	=
	(-1)^{n-1}  \sin \frac{n \pi}{2N}
	= (-1)^{n-1} Q_{1n} = (-1)^{n-1} R_n/\sqrt{N}.
	$}\fi%
$\sqrt{N} \bQ_{Nn} = (-1)^{n-1} R_n$ (which follows from basic trigonometry),
we can write
\[
\Vpu = \mathcal{C}\Avec,
\]
where
\begin{equation}
\mathcal{C} = \bmat{c_1 & \cdots & c_{N \shortminus 1}& c_\pi} \coloneqq \kappa_\text{pu} R \,
\scalebox{0.7}{$\bmat{(-1)^{N-1} & & & \\ & \ddots & & \\ & & -1 & \\ & & & 1}$},
\label{eq:def_C_matrix}
\end{equation}
and $\kappa_\text{pu}$ is a proportionality constant which can be assumed to be real (since the reference phase of $\Vpu$ can be chosen freely).

Combining \eqref{eq:mode_evolution_bb} with \eqref{eq:def_C_matrix} and also including cavity--beam interaction, quantified by parameters $\bm{\alpha}_n$ as in Section~\ref{sec:parasitic_modes}, we have the following state-space realization of the same-order-mode dynamics
\begin{subequations}%
\begin{align}
\dot{\Avec} &= \bm{\mathcal{A}} \Avec + \mathcal{B}_\text{g} \Fg + \bm{\mathcal{B}}_\text{b} \Ib \\
\Vpu &= \mathcal{C} \Avec
\label{eq:ss_model_baseband1}
\end{align}
where
\begin{align}
\bm{\mathcal{A}} &= -(\gamma_0 I + \Gamma_\ext) + i\Delta\Omega \\
\mathcal{B}_\text{g} &= \sqrt{2{\gamma_\ext}_\pi} R^\tran \\
\bm{B}_\text{b} &=
\frac{1}{2}
\bmat{
	\bm{\alpha}_1
	&
	\cdots &
	\bm{\alpha}_{N\text{-}1} &
	\alpha_\pi
}^\tran\\
\mathcal{C} &= \text{given by \eqref{eq:def_C_matrix}} \,.
\end{align}
\end{subequations}

\bibliography{same_order_modes_refs}

\begin{thebibliography}{14}%
\makeatletter
\providecommand \@ifxundefined [1]{%
 \@ifx{#1\undefined}
}%
\providecommand \@ifnum [1]{%
 \ifnum #1\expandafter \@firstoftwo
 \else \expandafter \@secondoftwo
 \fi
}%
\providecommand \@ifx [1]{%
 \ifx #1\expandafter \@firstoftwo
 \else \expandafter \@secondoftwo
 \fi
}%
\providecommand \natexlab [1]{#1}%
\providecommand \enquote  [1]{``#1''}%
\providecommand \bibnamefont  [1]{#1}%
\providecommand \bibfnamefont [1]{#1}%
\providecommand \citenamefont [1]{#1}%
\providecommand \href@noop [0]{\@secondoftwo}%
\providecommand \href [0]{\begingroup \@sanitize@url \@href}%
\providecommand \@href[1]{\@@startlink{#1}\@@href}%
\providecommand \@@href[1]{\endgroup#1\@@endlink}%
\providecommand \@sanitize@url [0]{\catcode `\\12\catcode `\$12\catcode
  `\&12\catcode `\#12\catcode `\^12\catcode `\_12\catcode `\%12\relax}%
\providecommand \@@startlink[1]{}%
\providecommand \@@endlink[0]{}%
\providecommand \url  [0]{\begingroup\@sanitize@url \@url }%
\providecommand \@url [1]{\endgroup\@href {#1}{\urlprefix }}%
\providecommand \urlprefix  [0]{URL }%
\providecommand \Eprint [0]{\href }%
\providecommand \doibase [0]{https://doi.org/}%
\providecommand \selectlanguage [0]{\@gobble}%
\providecommand \bibinfo  [0]{\@secondoftwo}%
\providecommand \bibfield  [0]{\@secondoftwo}%
\providecommand \translation [1]{[#1]}%
\providecommand \BibitemOpen [0]{}%
\providecommand \bibitemStop [0]{}%
\providecommand \bibitemNoStop [0]{.\EOS\space}%
\providecommand \EOS [0]{\spacefactor3000\relax}%
\providecommand \BibitemShut  [1]{\csname bibitem#1\endcsname}%
\let\auto@bib@innerbib\@empty
%</preamble>
\bibitem [{\citenamefont {Schilcher}(1998)}]{Schilcher1998}%
  \BibitemOpen
  \bibfield  {author} {\bibinfo {author} {\bibfnamefont {T.}~\bibnamefont
  {Schilcher}},\ }\emph {\bibinfo {title} {Vector Sum Control of Pulsed
  Accelerating Fields in {Lorentz} Force Detuned Superconducting Cavities}},\
  \href@noop {} {Ph.D. thesis},\ \bibinfo  {school} {University of Hamburg,
  Germany} (\bibinfo {year} {1998})\BibitemShut {NoStop}%
\bibitem [{\citenamefont {Ainsworth}\ and\ \citenamefont
  {Molloy}(2012)}]{Ainsworth2012}%
  \BibitemOpen
  \bibfield  {author} {\bibinfo {author} {\bibfnamefont {R.}~\bibnamefont
  {Ainsworth}}\ and\ \bibinfo {author} {\bibfnamefont {S.}~\bibnamefont
  {Molloy}},\ }\bibfield  {title} {\bibinfo {title} {Studies of parasitic
  cavity modes for proposed {ESS} linac lattices},\ }in\ \href@noop {} {\emph
  {\bibinfo {booktitle} {Proc. {LINAC2012}}}}\ (\bibinfo {year}
  {2012})\BibitemShut {NoStop}%
\bibitem [{\citenamefont {Ferrario}\ \emph {et~al.}(1996)\citenamefont
  {Ferrario}, \citenamefont {Mosnier}, \citenamefont {Serafini}, \citenamefont
  {Tazzioli},\ and\ \citenamefont {Tessier}}]{Ferrario1996}%
  \BibitemOpen
  \bibfield  {author} {\bibinfo {author} {\bibfnamefont {M.}~\bibnamefont
  {Ferrario}}, \bibinfo {author} {\bibfnamefont {A.}~\bibnamefont {Mosnier}},
  \bibinfo {author} {\bibfnamefont {L.}~\bibnamefont {Serafini}}, \bibinfo
  {author} {\bibfnamefont {F.}~\bibnamefont {Tazzioli}},\ and\ \bibinfo
  {author} {\bibfnamefont {J.}~\bibnamefont {Tessier}},\ }\bibfield  {title}
  {\bibinfo {title} {Multi-bunch energy spread induced by beam loading in a
  standing wave structure},\ }\href@noop {} {\bibfield  {journal} {\bibinfo
  {journal} {Part. Accel.}\ }\textbf {\bibinfo {volume} {52}},\ \bibinfo
  {pages} {1} (\bibinfo {year} {1996})}\BibitemShut {NoStop}%
\bibitem [{\citenamefont {Doolittle}(1989)}]{Doolittle1989}%
  \BibitemOpen
  \bibfield  {author} {\bibinfo {author} {\bibfnamefont {L.}~\bibnamefont
  {Doolittle}},\ }\href@noop {} {\emph {\bibinfo {title} {Understanding 5-cell
  mode structures}}},\ \bibinfo {type} {Tech. Rep.}\ \bibinfo {number}
  {CEBAF-TN-0120}\ (\bibinfo  {institution} {Jefferson Lab, Newport News, VA},\
  \bibinfo {year} {1989})\BibitemShut {NoStop}%
\bibitem [{\citenamefont {Padamsee}\ \emph {et~al.}(2008)\citenamefont
  {Padamsee}, \citenamefont {Knobloch},\ and\ \citenamefont
  {Hays}}]{Padamsee2008}%
  \BibitemOpen
  \bibfield  {author} {\bibinfo {author} {\bibfnamefont {H.}~\bibnamefont
  {Padamsee}}, \bibinfo {author} {\bibfnamefont {J.}~\bibnamefont {Knobloch}},\
  and\ \bibinfo {author} {\bibfnamefont {T.}~\bibnamefont {Hays}},\ }\href@noop
  {} {\emph {\bibinfo {title} {{RF} Superconductivity for Accelerators}}},\
  \bibinfo {edition} {2nd}\ ed.\ (\bibinfo  {publisher} {Wiley-VCH},\ \bibinfo
  {address} {Weinheim, Germany},\ \bibinfo {year} {2008})\BibitemShut {NoStop}%
\bibitem [{\citenamefont {Wangler}(2008)}]{Wangler2008}%
  \BibitemOpen
  \bibfield  {author} {\bibinfo {author} {\bibfnamefont {T.~P.}\ \bibnamefont
  {Wangler}},\ }\href@noop {} {\emph {\bibinfo {title} {RF Linear
  Accelerators}}},\ \bibinfo {edition} {2nd}\ ed.\ (\bibinfo  {publisher}
  {Wiley-VCH},\ \bibinfo {address} {Weinheim, Germany},\ \bibinfo {year}
  {2008})\BibitemShut {NoStop}%
\bibitem [{\citenamefont {Sekutowicz}(2010)}]{Sekutowicz2010}%
  \BibitemOpen
  \bibfield  {author} {\bibinfo {author} {\bibfnamefont {J.}~\bibnamefont
  {Sekutowicz}},\ }\bibinfo {title} {Superconducting elliptical cavities},\ in\
  \href@noop {} {\emph {\bibinfo {booktitle} {Proc. {CERN} Accel. School ---
  {RF} for Accelerators}}}\ (\bibinfo  {publisher} {CERN},\ \bibinfo {address}
  {Geneva, Switzerland},\ \bibinfo {year} {2010})\BibitemShut {NoStop}%
\bibitem [{\citenamefont {Troeng}(2019)}]{Troeng2019Thesis}%
  \BibitemOpen
  \bibfield  {author} {\bibinfo {author} {\bibfnamefont {O.}~\bibnamefont
  {Troeng}},\ }\emph {\bibinfo {title} {Cavity Field Control for Linear
  Particle Accelerators}},\ \href@noop {} {Ph.D. thesis},\ \bibinfo  {school}
  {Lund University, Sweden} (\bibinfo {year} {2019})\BibitemShut {NoStop}%
\bibitem [{\citenamefont {Troeng}(2020)}]{Troeng2020CavityModel}%
  \BibitemOpen
  \bibfield  {author} {\bibinfo {author} {\bibfnamefont {O.}~\bibnamefont
  {Troeng}},\ }\href@noop {} {\bibinfo {title} {Energy-based parameterization
  of accelerating-mode dynamics}} (\bibinfo {year} {2020}),\ \Eprint
  {https://arxiv.org/abs/2009.14813} {arXiv:2009.14813 [physics.acc-ph]}
  \BibitemShut {NoStop}%
\bibitem [{\citenamefont {Vogel}(2007)}]{Vogel2007}%
  \BibitemOpen
  \bibfield  {author} {\bibinfo {author} {\bibfnamefont {E.}~\bibnamefont
  {Vogel}},\ }\bibfield  {title} {\bibinfo {title} {High gain proportional {RF}
  control stability at {TESLA} cavities},\ }\href@noop {} {\bibfield  {journal}
  {\bibinfo  {journal} {Phys. Rev. Accel. Beams}\ }\textbf {\bibinfo {volume}
  {10}},\ \bibinfo {pages} {052001} (\bibinfo {year} {2007})}\BibitemShut
  {NoStop}%
\bibitem [{\citenamefont {Liepe}(2001)}]{Liepe2001}%
  \BibitemOpen
  \bibfield  {author} {\bibinfo {author} {\bibfnamefont {M.~U.}\ \bibnamefont
  {Liepe}},\ }\emph {\bibinfo {title} {Superconducting Multicell Cavities for
  Linear Colliders}},\ \href@noop {} {Ph.D. thesis},\ \bibinfo  {school}
  {University of Hamburg, Germany} (\bibinfo {year} {2001})\BibitemShut
  {NoStop}%
\bibitem [{\citenamefont {Zhou}\ \emph {et~al.}(1996)\citenamefont {Zhou},
  \citenamefont {Doyle},\ and\ \citenamefont {Glover}}]{Zhou1996}%
  \BibitemOpen
  \bibfield  {author} {\bibinfo {author} {\bibfnamefont {K.}~\bibnamefont
  {Zhou}}, \bibinfo {author} {\bibfnamefont {J.~C.}\ \bibnamefont {Doyle}},\
  and\ \bibinfo {author} {\bibfnamefont {K.}~\bibnamefont {Glover}},\
  }\href@noop {} {\emph {\bibinfo {title} {Robust and Optimal Control}}}\
  (\bibinfo  {publisher} {Prentice Hall},\ \bibinfo {address} {Englewood
  Cliffs, NJ},\ \bibinfo {year} {1996})\BibitemShut {NoStop}%
\bibitem [{\citenamefont {Skogestad}\ and\ \citenamefont
  {Postlethwaite}(2007)}]{Skogestad2007}%
  \BibitemOpen
  \bibfield  {author} {\bibinfo {author} {\bibfnamefont {S.}~\bibnamefont
  {Skogestad}}\ and\ \bibinfo {author} {\bibfnamefont {I.}~\bibnamefont
  {Postlethwaite}},\ }\href@noop {} {\emph {\bibinfo {title} {Multivariable
  Feedback Control: Analysis and Design}}},\ \bibinfo {edition} {2nd}\ ed.\
  (\bibinfo  {publisher} {John Wiley \& Sons},\ \bibinfo {address} {Chichester,
  UK},\ \bibinfo {year} {2007})\BibitemShut {NoStop}%
\bibitem [{\citenamefont {Haus}(1983)}]{Haus1983}%
  \BibitemOpen
  \bibfield  {author} {\bibinfo {author} {\bibfnamefont {H.~A.}\ \bibnamefont
  {Haus}},\ }\href@noop {} {\emph {\bibinfo {title} {Waves and Fields in
  Optoelectronics}}}\ (\bibinfo  {publisher} {Prentice-Hall},\ \bibinfo
  {address} {Englewood Cliffs, NJ},\ \bibinfo {year} {1983})\BibitemShut
  {NoStop}%
\end{thebibliography}%

\onecolumngrid % RevTeX specific way to achieve balancing

\end{document}